\documentclass[apj]{emulateapj}

\usepackage{apjfonts,ulem,hyperref}
\usepackage{amssymb, amsmath}
\newcommand\degree{\degr}
\newcommand\degrees\degree

\DeclareSymbolFont{UPM}{U}{eur}{m}{n}
\DeclareMathSymbol{\umu}{0}{UPM}{"16}
\let\oldumu=\umu
\renewcommand\umu{\ifmmode\oldumu\else\math{\oldumu}\fi}
\newcommand\micro{\umu}
\renewcommand\micron{\micro m}
\newcommand\microns \micron

\let\oldsim=\sim
\renewcommand\sim{\ifmmode\oldsim\else\math{\oldsim}\fi}
\let\oldpm=\pm
\renewcommand\pm{\ifmmode\oldpm\else\math{\oldpm}\fi}
\newcommand\by{\ifmmode\times\else\math{\times}\fi}

\newbox{\wdbox}
\renewcommand\c{\setbox\wdbox=\hbox{,}\hspace{\wd\wdbox}}
\renewcommand\i{\setbox\wdbox=\hbox{i}\hspace{\wd\wdbox}}

\newcount\timect
\newcount\hourct
\newcount\minct
\newcommand\now{\timect=\time \divide\timect by 60
         \hourct=\timect \multiply\hourct by 60
         \minct=\time \advance\minct by -\hourct
         \number\timect:\ifnum \minct < 10 0\fi\number\minct}

\catcode`@=11

\newcommand\comment[1]{}
\newcommand\commenton{\catcode`\%=14}
\newcommand\commentoff{\catcode`\%=12}

\renewcommand\math[1]{$#1$}
\newcommand\mathshifton{\catcode`\$=3}
\newcommand\mathshiftoff{\catcode`\$=12}

\comment{the backslash is necessary}

\comment{alignment tab}

\let\atab=&
\newcommand\atabon{\catcode`\&=4}
\newcommand\ataboff{\catcode`\&=12}

\let\oldmsp=\sp
\let\oldmsb=\sb
\def\sp#1{\ifmmode
           \oldmsp{#1}%
         \else\strut\raise.85ex\hbox{\scriptsize #1}\fi}
\def\sb#1{\ifmmode
           \oldmsb{#1}%
         \else\strut\raise-.54ex\hbox{\scriptsize #1}\fi}
\newbox\@sp
\newbox\@sb
\def\sbp#1#2{\ifmmode%
           \oldmsb{#1}\oldmsp{#2}%
         \else
           \setbox\@sb=\hbox{\sb{#1}}%
           \setbox\@sp=\hbox{\sp{#2}}%
           \rlap{\copy\@sb}\copy\@sp
           \ifdim \wd\@sb >\wd\@sp
             \hskip -\wd\@sp \hskip \wd\@sb
           \fi
        \fi}
\def\msp#1{\ifmmode
           \oldmsp{#1}
         \else \math{\oldmsp{#1}}\fi}
\def\msb#1{\ifmmode
           \oldmsb{#1}
         \else \math{\oldmsb{#1}}\fi}
\def\supon{\catcode`\^=7}
\def\supoff{\catcode`\^=12}
\def\subon{\catcode`\_=8}
\def\suboff{\catcode`\_=12}
\def\supsubon{\supon \subon}
\def\supsuboff{\supoff \suboff}

\newcommand\actcharon{\catcode`\~=13}
\newcommand\actcharoff{\catcode`\~=12}

\newcommand\paramon{\catcode`\#=6}
\newcommand\paramoff{\catcode`\#=12}

\comment{And now to turn us totally on and off...}

\newcommand\reservedcharson{\commenton \mathshifton \atabon \supsubon \actcharon
	\paramon}

\newcommand\reservedcharsoff{\commentoff \mathshiftoff \ataboff
	\supsuboff \actcharoff \paramoff}

\catcode`@=12
\reservedcharsoff

\reservedcharson

\comment{ Must have ONLY ONE of these... trust these macros, they work

}

\newcommand{\squishlist}{
 \begin{list}{$\bullet$}
  { \setlength{\itemsep}{1pt}
     \setlength{\parsep}{0pt}
     \setlength{\topsep}{3pt}
     \setlength{\partopsep}{0pt}
     \setlength{\leftmargin}{2.0em}
     \setlength{\labelwidth}{1.5em}
     \setlength{\labelsep}{0.5em} } }

\newcommand{\squishend}{
  \end{list}  }

\reservedcharson
\actcharon
\shorttitle{Quantifying and Predicting the Presence of Clouds in Exoplanet Atmospheres}
\shortauthors{K. B. Stevenson}

\submitted{}

\begin{document}

\title{Quantifying and Predicting the Presence of Clouds in Exoplanet Atmospheres}

\author{Kevin B.\ Stevenson\altaffilmark{1,2}}
\affil{\sp{1}Department of Astronomy and Astrophysics, University of Chicago, 5640 S Ellis Ave, Chicago, IL 60637, USA}
\affil{\sp{2}NASA Sagan Fellow}

\email{E-mail: kbs@uchicago.edu}

\begin{abstract}
One of the most outstanding issues in exoplanet characterization is understanding the prevalence of obscuring clouds and hazes in their atmospheres.  The ability to predict the presence of clouds/hazes {\em a priori} is an important goal when faced with limited telescope resources and advancements in atmospheric characterization that rely on the detection of spectroscopic features.  As a means to identify favorable targets for future studies with {\em HST} and {\em JWST}, we use published {\em HST}/WFC3 transmission spectra to determine the strength of each planet's water feature, as defined by the H\sb{2}O~--~J index.  By expressing this parameter in units of atmospheric scale height, we provide a means to efficiently compare the size of spectral features over a physically diverse sample of exoplanets.  We find the H\sb{2}O~--~J index to be strongly correlated with planet temperature when $T$\sb{eq}$<750^{+90}_{-60}$~K and weakly correlated with surface gravity for planets with $\log g < 3.2^{+0.3}_{-0.2}$~dex.  Otherwise, the median value of the H\sb{2}O~--~J index is $1.8{\pm}0.3$~H.  Using these two physical parameters, we identify a division between ``classes'' of exoplanets, such that objects above $T$\sb{eq}$=700$~K and $\log g = 2.8$~dex are more likely to have clearer atmospheres with stronger spectral features (H\sb{2}O~--~J > 1) and those below at least one of these thresholds are increasingly likely to have predominantly cloudy atmospheres with muted spectral features (H\sb{2}O~--~J < 1).  Additional high-precision measurements are needed to corroborate the reported trends.
\end{abstract}

\section{Introduction}
\label{intro}

In the last few years, there has been a surge of publications reporting reliable constraints on the transmission spectra of extrasolar planets.  Using {\em HST}/WFC3, many of these publications have reported the detection of H\sb{2}O, thus revealing insight into their chemical makeup.  However, a substantial fraction have also reported statistically flat transmission spectra, which inhibit further characterization and understanding.  The strength of the measured water feature can be reduced by the presence of obscuring clouds/hazes, high metallicities causing high mean molecular weight atmospheres, and/or an extremely low H\sb{2}O abundance \citep[e.g., ][]{Seager2000, Charbonneau2002, MillerRicci2009, Morley2013, Moses2013b, Madhu2014}.  However, the production of clouds and hazes is generally believed to be the most common mechanism for causing muted spectroscopic features in exoplanet spectra that have been obtained so far \citep{Sing2015b}.

It is important to understand why some planets have predominantly cloud-free atmospheres, thus revealing insight into their compositions, while others appear to contain high-altitude clouds or hazes that obscure our measurements.  The ability to predict the presence of obscuring clouds/hazes {\em a priori} can increase telescope efficiency by allowing future investigations to concentrate on targets most likely to contain relatively clear atmospheres and strong spectroscopic signals.  This is especially true for the {\em James Webb Space Telescope (JWST)}, which is scheduled to launch in 2018 \citep{Beichman2014,Barstow2015,Cowan2015}.

The goal of this investigation is to use exoplanets' physical properties to predict the strength of spectroscopic features (thus inferring the presence of clouds/hazes) in the near-infrared.  For example, the correlation between clouds and temperature is well known in brown dwarfs \citep[e.g., ][]{Ackerman2001,Lodders2002,Burgasser2006b}.  There is also evidence for a cloud-gravity dependence near the L-T transition \citep[e.g., ][]{Burrows2006,Metchev2006,Saumon2008}. Using high-precision, space-based transmission spectra, we look for similar correlations in our sample exoplanet population.

The layout of this paper is as follows.  Section \ref{sec:proc} describes our procedure for quantifying the strength of the water feature using the H\sb{2}O~--~J index.  In Section \ref{sec:results}, we investigate possible trends between our sample planet population's reported H\sb{2}O~--~J index values and their physical properties.  Finally, Section \ref{sec:future} discusses what future observations are needed to refine our constraints and further our understanding of exoplanet atmospheres.

\section{PROCEDURE}
\label{sec:proc}

In taking our first step, we look to select a sample population of transiting exoplanets with well-characterized atmospheres.  Although there is a significant cumulation of {\em Spitzer Space Telescope} transit data and increasing efforts to obtain ground-based transmission spectra of exoplanets, for this investigation we limit our sample population to only include {\em HST}/WFC3 data acquired with the G141 grism.  These high-precision spectra have been shown to reliably and repeatably resolve the water feature centered at 1.4 {\microns} without confusion from other absorption features or variability caused by Earth's atmosphere.  Furthermore, the {\em HST}/WFC3 sample population is both significant in size and diverse in their physical properties.

\begin{table*}[tb]
\centering
\caption{\label{tab:planetparams} 
Planet Parameters And Sources}
\begin{tabular}{ccccccl}
    \hline
    \hline      
    Planet      & T\sb{eq}  & Log(g)    & H\sb{2}O~--~J & H\sb{2}O      & J             & Reference\tablenotemark{a} \\
                & (K)       & (cgs)     & (H)           & ({\microns})  & ({\microns})  &           \\
    \hline
    GJ 436b     & 649       & 3.10      & 0.54{\pm}0.46 & 1.230--1.289  & 1.362--1.438  & \citet{Knutson2014}       \\
    GJ 1214b    & 559       & 2.94      & 0.02{\pm}0.10 & 1.228--1.297  & 1.366--1.435  & \citet{Kreidberg2014}     \\
    HAT-P-1b    & 1303      & 2.93      & 2.13{\pm}0.61 & 1.223--1.300  & 1.376--1.434  & \citet{Wakeford2013}      \\
    HAT-P-11b   & 870       & 3.06      & 3.47{\pm}0.72 & 1.228--1.303  & 1.360--1.435  & \citet{Fraine2014}        \\
    HAT-P-12b   & 957       & 2.75      & 0.21{\pm}0.60 & 1.226--1.297  & 1.367--1.438  & \citet{Line2013c}         \\
    HD 97658b   & 733       & 3.15      & 1.85{\pm}0.91 & 1.237--1.292  & 1.366--1.440  & \citet{Knutson2014b}      \\
    HD 189733b  & 1199      & 3.34      & 1.86{\pm}0.36 & 1.222--1.297  & 1.372--1.447  & \citet{McCullough2014}    \\
    HD 209458b  & 1445      & 2.97      & 1.15{\pm}0.13 & 1.232--1.288  & 1.364--1.439  & \citet{Deming2013}        \\
    WASP-12b    & 2581      & 3.02      & 1.48{\pm}0.32 & 1.251--1.320  & 1.389--1.458  & \citet{Kreidberg2015}     \\
    WASP-17b    & 1547      & 2.53      & 0.67{\pm}0.29 & 1.240--1.296  & 1.381--1.437  & \citet{Mandell2013}       \\
    WASP-19b    & 2064      & 3.16      & 3.03{\pm}0.64 & 1.230--1.286  & 1.371--1.427  & \citet{Mandell2013}       \\
    WASP-31b    & 1573      & 2.70      & 0.86{\pm}0.48 & 1.234--1.294  & 1.374--1.434  & \citet{Sing2015}          \\
    WASP-43b    & 1374      & 3.70      & 1.08{\pm}0.50 & 1.228--1.297  & 1.366--1.435  & \citet{Kreidberg2014b}    \\
    XO-1b       & 1206      & 3.19      & 1.66{\pm}0.55 & 1.234--1.290  & 1.365--1.422  & \citet{Deming2013}        \\
    \hline
\end{tabular}
\tablenotetext{1}{The references provide the measured transit depths or radius ratios from which we calculate the H\sb{2}O~--~J index within the specified wavelength regions.}
\end{table*}

\begin{figure}[tb]
\centering
\includegraphics[width=1.0\linewidth]{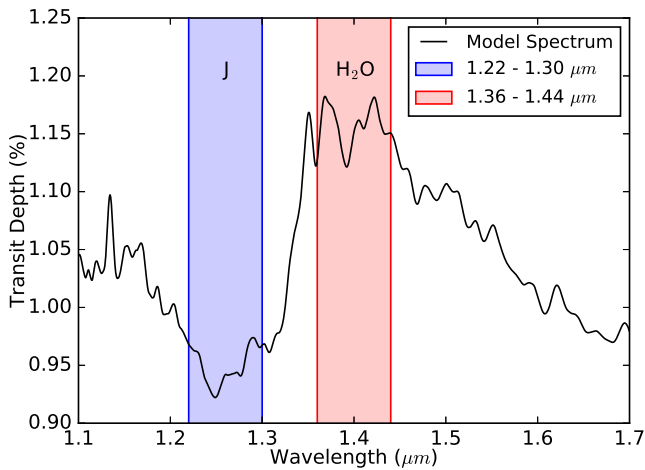}
\caption{\label{fig:h2o-j}{
Defined wavelength regions to estimate the strength of a planet's water feature.  For each planet with a published {\em HST}/WFC3 transmission spectrum, we compute the difference in transit depth between these two spectral regions and convert to units of atmospheric scale height.  This defines the H\sb{2}O~--~J index.
}}
\end{figure}

Next, we need to establish wavelength regions both inside and outside of our selected spectroscopic feature.  We define two spectral regions: (1) the peak of the broad water feature from 1.36 to 1.44 {\microns} and (2) a J-band baseline from 1.22 to 1.30 {\microns} (see Figure \ref{fig:h2o-j}).  Compared to brown dwarf measurements, exoplanet transmission spectra have a higher degree of uncertainty; therefore, we have chosen relatively wide spectral regions to improve the precision of our comparisons.  Also, authors report transit depths or radius ratios using spectrophotometric bins of varying widths; therefore, utilizing wide spectral windows minimizes the effects of having slightly offset spectral regions for different planets (see Table \ref{tab:planetparams}).  Establishing new transit depths over our specified spectral regions would require reanalyzing all of the available {\em HST}/WFC3 data and is outside the scope of this investigation.  Future analyses, however, are encouraged to report transit depths and uncertainties over these two spectral regions.  As a cautionary note, using a dual-band measure of the water feature strength has its limitations.  Inherently noisy spectra \citep[e.g., HD~97658b, ][]{Knutson2014b} may not exhibit distinct water features, but can still produce H\sb{2}O~--~J index values > 1 if the measured transit depths exhibit wavelength-dependent trends.

For our third step, we are concerned with establishing a universal metric to which planets with varying physical properties can be compared.  For this, we adopt the atmospheric scale height, 
\begin{equation}
H = \frac{k\sb{B}T\sb{eq}}{\mu g},
\end{equation}
\noindent where $k\sb{B}$ is Boltzmann's constant, $T$\sb{eq} is the planet's equilibrium temperature, $\mu$ is the atmospheric mean molecular weight, and $g$ is the planet's surface gravity (in MKS units).  For planets with $R>0.9$ Jupiter-radii, we adopt Jupiter's atmospheric mean molecular weight ($\mu$ = 2.2~u).  For planets with $R<0.5$ Jupiter-radii, we compute an atmospheric mean molecular weight of 3.8~u using the prescription of \citet{Fortney2013} and assuming 80$\times$Solar metallicity, which is comparable to both Uranus and Neptune \citep{Kreidberg2014b}.  All of the planets in our sample are expected to have H\sb{2}/He-dominated atmospheres \citep{Rogers2015}.  One advantage of adopting the scale height as our metric is that it accounts for differences in our sample population's equilibrium temperatures and surface gravities, two prominent factors thought to affect the production of clouds and hazes in exoplanet atmospheres.
 
For each planet, we compute the scale height using the $T$\sb{eq} and $\log g$ values listed in Table \ref{tab:planetparams}, which we derive from data provided by the Exoplanet Orbit Database \citep[www.exoplanets.org, ][]{Han2014}.  Our $T$\sb{eq} calculation assumes zero albedo and full heat redistribution.  For ease of calculation, we convert the scale height of each planet into a change in transit depth, $\Delta D$, using the following relation:
\begin{equation}
\Delta D \sim \frac{2HR_P}{R_S^2},
\end{equation}
\noindent where $R$\sb{P} and $R$\sb{S} are the planet and stellar radii, respectively.  Therefore, $\Delta D$ is the change in transit depth that corresponds to a one-scale-height change in altitude and serves as a normalizing factor between planetary systems with different physical properties.  Next, we compute the difference in transit depth between our two spectral regions, ensuring to propagate uncertainties, then divide by $\Delta D$ for each planet.  This defines the H\sb{2}O~--~J index, which quantifies the fractional change in atmospheric scale height between our two spectral regions.  For the purposes of this investigation, we discard planets whose H\sb{2}O~--~J index uncertainties are larger than one scale height; this includes GJ~3470b, TrES-2b, TrES-4b, and CoRoT-1b.  Due to their large uncertainties, these measurements are uninformative and only serve to obfuscate trends in certain figures.  Table \ref{tab:planetparams} lists the remaining H\sb{2}O~--~J index values in units of scale height.

Finally, to allow for a more intuitive understanding, we qualitatively describe the state of a planet's atmosphere assuming that changes in the H\sb{2}O~--~J index value are solely a function of cloud cover.  Planets with the largest H\sb{2}O~--~J index values ($>$ 2) are likely to have cloud-free atmospheres at the depths probed by the WFC3 measurements.  Those with values between one and two may be thought of as having partially cloudy to mostly clear skies, or a thin haze layer.  Planet atmospheres with H\sb{2}O~--~J index values less than unity have the weakest water features (if any) that are likely caused by increased cloud cover or haze thickness at or above the probed depths.

\section{ANALYSES AND RESULTS}
\label{sec:results}

\begin{figure}[tb]
\centering
\includegraphics[width=1.0\linewidth]{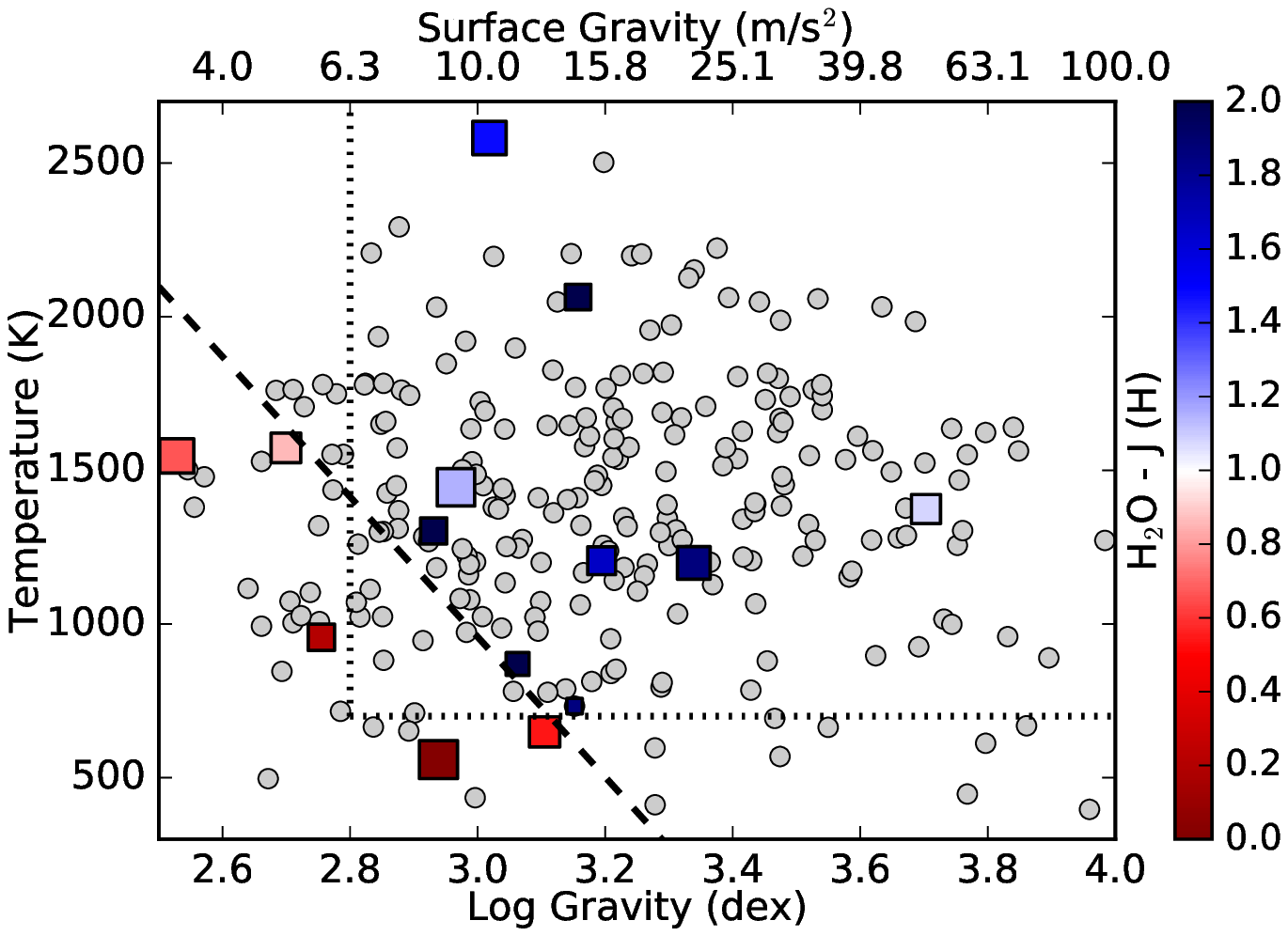}
\includegraphics[width=1.0\linewidth]{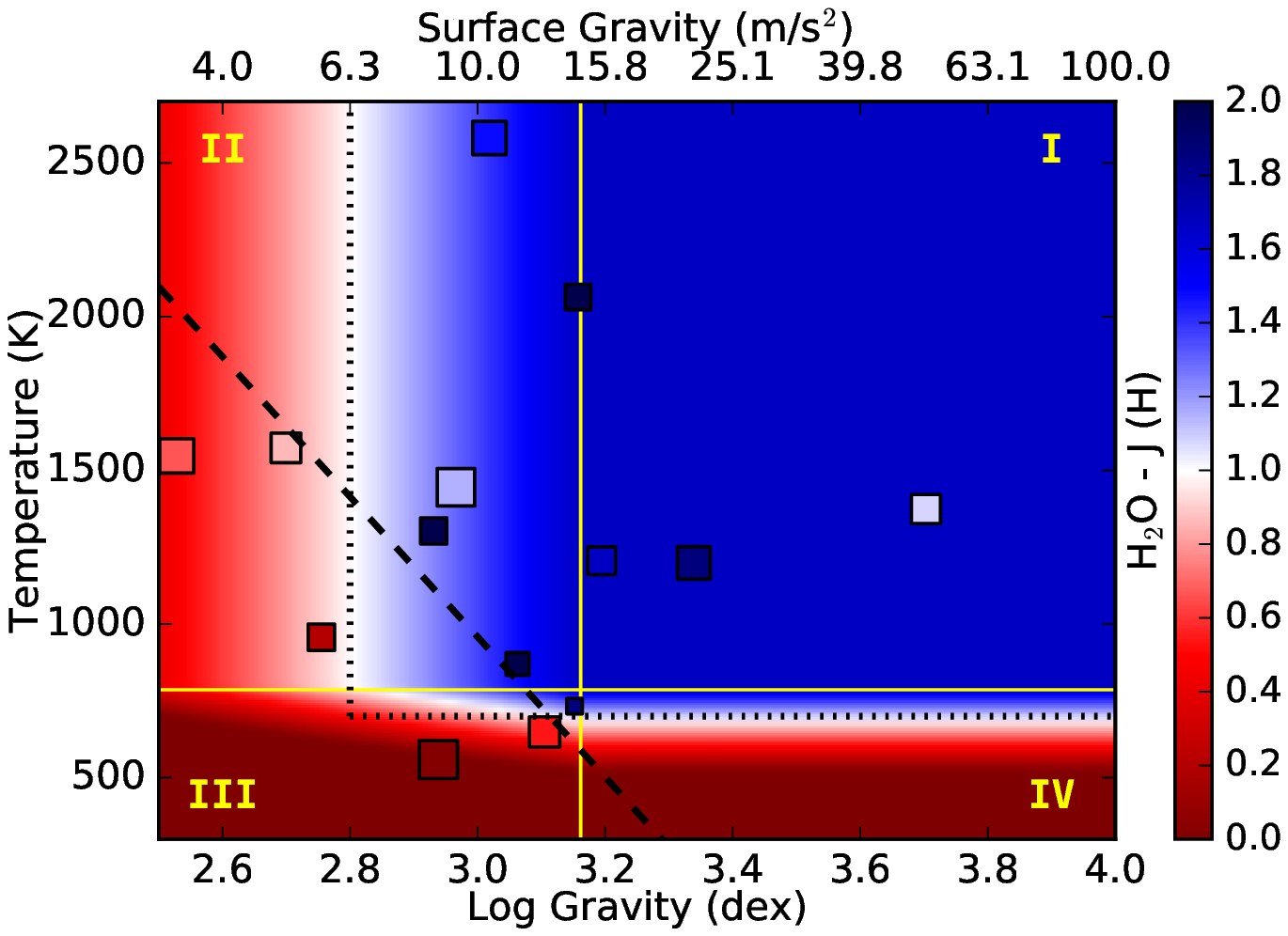}
\caption{\label{fig:TeqLogg}{
Water feature strength as a function of equilibrium temperature and surface gravity.  The nine planets with strong measured water features (H\sb{2}O~--~J $>$ 1, blue squares) have $T$\sb{eq} $> 700$~K and $\log g > 2.8$ (dotted lines).  The remaining five planets with H\sb{2}O~--~J $<$ 1 (red squares) have muted water features.  The size of the squares is inversely proportional to the uncertainty in the H\sb{2}O~--~J index, thus larger symbols carry more weight.  Like the dotted lines, the dashed diagonal line also delineates these two ``classes'' of exoplanets.   Within the top panel, the gray circles depict confirmed transiting exoplanets with known masses and radii.  The coloring in the bottom panel depicts the best-fit 2D model segmented into four quadrants (yellow lines and numerals).  Additional measurements are needed to more precisely determine the transition region for atmospheres with and without obscuring clouds/hazes at the depths probed by WFC3.
}}
\end{figure}

After exploring how the water-feature strength varies as a function of the planets' physical properties (i.e. mass, radius, temperature, rotation rate, etc.), we find that the H\sb{2}O~--~J index is strongly correlated with equilibrium temperature and weakly correlated with surface gravity (typically expressed as $\log g$).  The top panel of Figure \ref{fig:TeqLogg} illustrates this dependence by dividing the sample population into two ``classes'' of exoplanets using H\sb{2}O~--~J~=~1 as a borderline.  Those with H\sb{2}O~--~J $>$ 1 are bounded in phase space by $T$\sb{eq} $> 700$~K and $\log g > 2.8$.  Recall that for these planets, any putative atmospheric clouds would play only a minor role in altering the strength of the measured water features.  Conversely, for the sample population with H\sb{2}O~--~J $<$ 1, clouds and/or hazes likely play an important role in muting their spectroscopic features.

In the bottom panel of Figure \ref{fig:TeqLogg}, we include a best-fit 2D model that simultaneously fits the temperature and gravity dependence.  This model consists of four quadrants separated by freely-varying transition (or knot) values.  Above the temperature and gravity knot values (Quadrant I), we adopt a constant H\sb{2}O~--~J index value (one free parameter, $c$\sb{I}) because the available data exhibit no significant trend.  Quadrants II and IV assume linear dependences in gravity and temperature, respectively (two free parameters each, $m$\sb{II}, $b$\sb{II}, $m$\sb{IV}, and $b$\sb{IV}).  To complete the model, we apply bilinear interpolation in Quadrant III.  The final model contains five free parameters and the knot values are computed at each step of our Markov-Chain Monte Carlo analysis \citep[MCMC, ][]{Ford2005}.

We derive the following median parameter values with 1$\sigma$ uncertainties: $c$\sb{I} = $1.8{\pm}0.3$~H, $m$\sb{II} = $1.9{\pm}0.7$~H/dex, $b$\sb{II} = $-4.3{\pm}1.9$~H, $m$\sb{IV} = $0.0074^{+0.0034}_{-0.0024}$~H/K, and $b$\sb{IV} = $-3.7^{+1.4}_{-1.8}$~H.  Interestingly, Quadrant IV from the best-fit model does not contain any planets from our sample, but the free parameters are tightly constrained by the population in Quadrant III.  Nonetheless, the model would benefit from having {\em HST}/WFC3 measurements within this region.  We derive a median transition temperature of $750^{+90}_{-60}$~K and a median transition $\log g$ of $3.2^{+0.3}_{-0.2}$~dex.  With 14 data points and five free parameters, our best-fit model achieves a $\chi^2$ value of 19.8.

We test our proposed temperature dependence by computing the variance-weighted experimental linear-correlation coefficient, $r$, which ranges from -1 to 1 \citep{Bevington2003}.  The three planets in Quadrant III have $r = 0.96$, indicating a near-total positive correlation between planet temperature and water feature size, and a p-value of 0.092.  Conversely, the planets in Quadrants I and II have $r = 0.06$, which is consistent with no correlation.  Next, we repeat the test using surface gravity as the correlating parameter.  The eight planets in Quadrant II exhibit a moderate positive correlation ($r = 0.61$) with a relatively small p-value (0.053).  The three planets in Quadrant I have a surprisingly strong negative correlation ($r = -0.85$) with surface gravity, but they also have large uncertainties on their H\sb{2}O~--~J index values and a large p-value (0.18).  So, we repeat this test while including three additional planets from Quadrant II with $\log g > 3.0$ (the 1$\sigma$ lower limit on the transition surface gravity).  In this case, we find a much weaker correlation ($r = -0.32$) that suggests a larger sample size is needed in Quadrant I before any conclusions can be drawn.

\subsection{Temperature Dependence}
\label{sec:temp}

\begin{figure}[tb]
\centering
\includegraphics[width=1.0\linewidth]{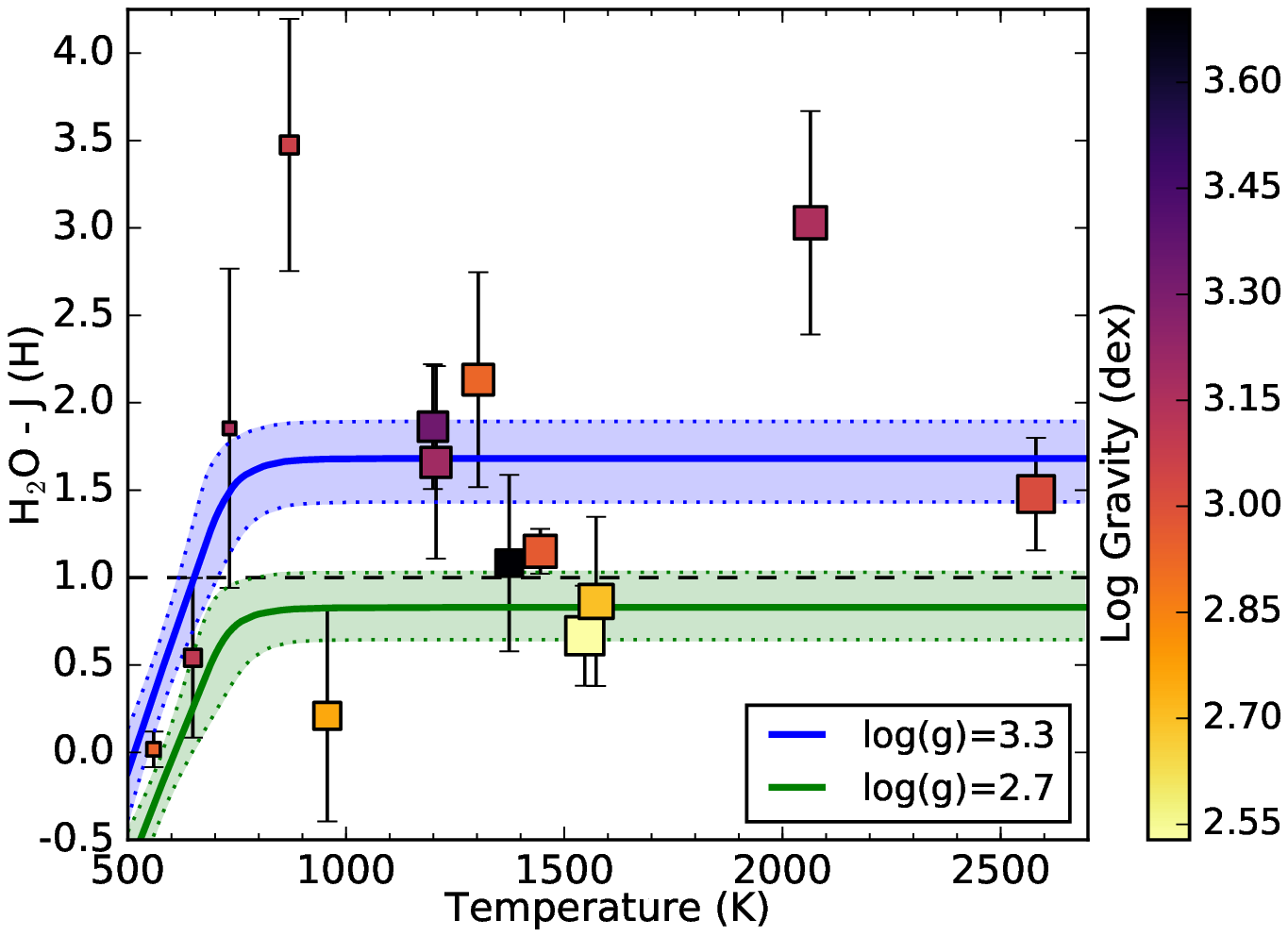}
\includegraphics[width=1.0\linewidth]{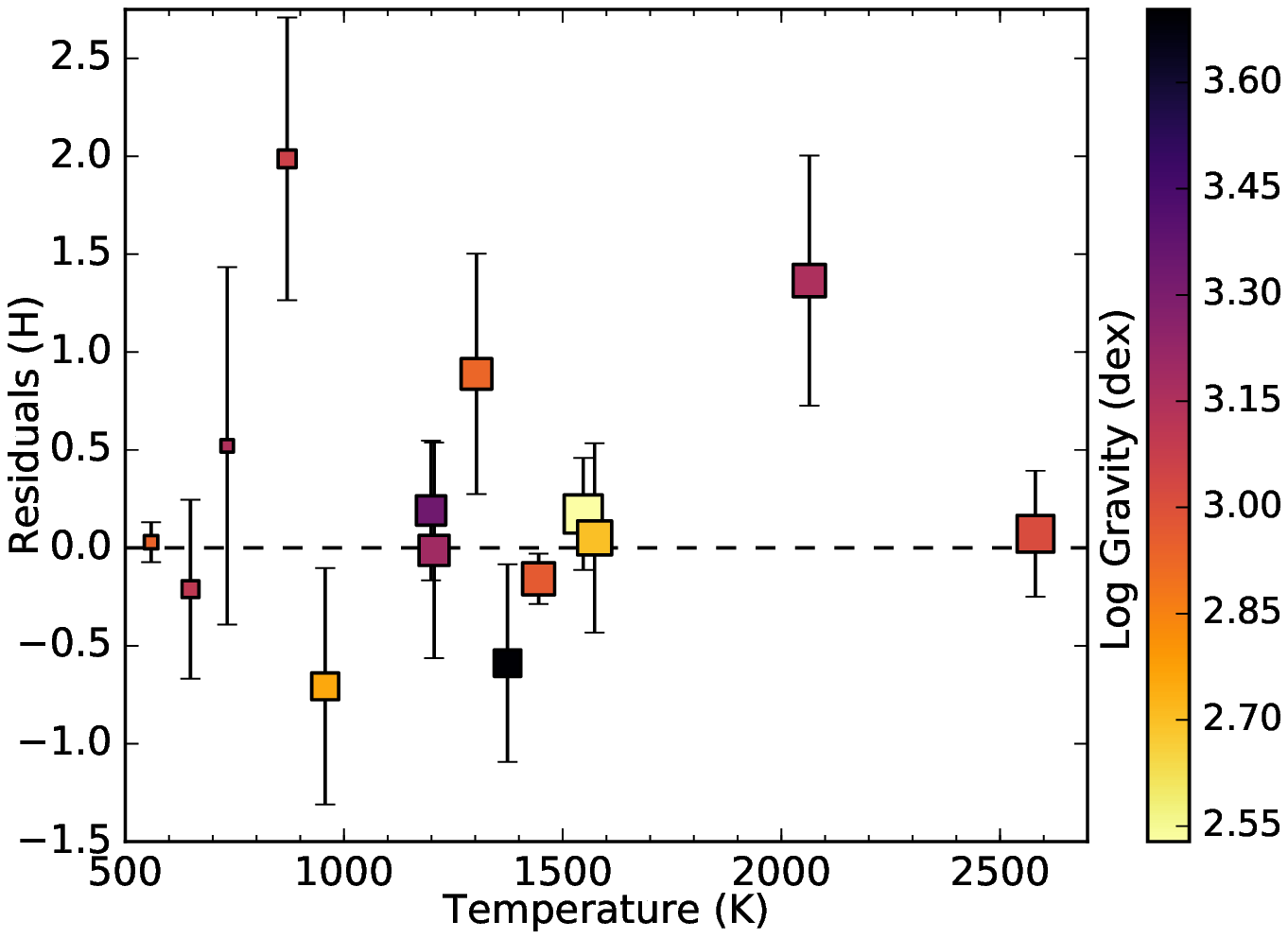}
\caption{\label{fig:TeqDep}{
H\sb{2}O~--~J index dependence on planet temperature.  The symbol size is proportional to planet radius and symbol color is proportional to $\log g$.  The top panel depicts the water feature strength with 1$\sigma$ uncertainties.  The blue and green lines with colored 1$\sigma$ uncertainty regions are representative models at two different surface gravities ($\log g = 3.2$ and 2.7~dex, respectively).  Curvature in the median model and uncertainties near 800 K results from allowing the knot value to vary in our fits.  The bottom panel shows no discernible correlation between the residuals and planet temperature.}}
\end{figure}

Here we further examine the H\sb{2}O~--~J index dependence on planet temperature.  The top panel of Figure \ref{fig:TeqDep} plots the water feature strength (with uncertainties) versus equilibrium temperature for all 14 exoplanets in our sample population.  From this, it is clear that the model is strongly influenced by the relatively small uncertainties of GJ~1214b and HD~209458b.  Fitting a linear trend to the entire dataset (not shown) results in a 8$\sigma$ detection of a slope, thus suggesting a strong temperature dependence at face value.  However, the quality of the fit is relatively poor ($\chi_{\nu}^2 = 4.7$), thus justifying the use of our more complex model (discussed above).  With an improved fit ($\chi_{\nu}^2 = 2.5$), we report a correlation between the water feature size and planet temperature at 3.1$\sigma$ confidence when $T$\sb{eq}~<~750~K.

Next, we consider how planet radius might affect the water feature size because the four coolest planets are also the smallest.  The sub-Neptune-size planet HD~97658b (2.25~$R_{\oplus}$) is smaller than GJ 1214b (2.68~$R_{\oplus}$), but it has a larger H\sb{2}O~--~J index value (1.85{\pm}0.91 vs.~0.02{\pm}0.10).  Furthermore, HAT-P-11b is comparable in size to GJ 436b but has the largest H\sb{2}O~--~J index value (3.47{\pm}0.72 vs.~0.54{\pm}0.46).  These examples argue against a strong correlation between the H\sb{2}O~--~J index value and planet radius.  It may be relevant that the two coolest planets both orbit M dwarf stars because their intense UV radiation may facilitate the production of obscuring aerosols.  Therefore, as a critical next step, we recommend additional high-precision {\em HST}/WFC3 measurements of Jupiter-size exoplanets with $T$\sb{eq} = 500 -- 700~K and orbiting F, G, or K stars to validate the reported temperature dependence.

The bottom panel of Figure \ref{fig:TeqDep} plots the H\sb{2}O~--~J index residuals from our best-fit model versus planet temperature.  We note that 57\% of our sample population are within 1$\sigma$ of zero and 86\% are within 2$\sigma$.  The scatter in the residuals may be reduced by applying more sophisticated models or examining correlations between the H\sb{2}O~--~J index and other physical parameters not considered in this analysis.  Most WFC3 light curves do not exhibit time-correlated (red) noise and model fits tend to be repeatable and near the photon limit; therefore, the H\sb{2}O~--~J index uncertainties are not likely to be significantly underestimated.

The temperature dependence of our best-fit model, including the $750^{+90}_{-60}$~K transition temperature, may be explained by the production of photochemical hazes \citep{Morley2015}.  At temperatures below $\sim$1100~K, CH\sb{4} becomes the dominant carbon-bearing molecule for an atmosphere in thermochemical equilibrium.  Methane photochemistry, which is initiated by the absorption of incident UV radiation, results in the formation of soot precursors such as acetylene (C\sb{2}H\sb{2}) and hydrogen cyanide (HCN) in the upper atmospheric layers that can then lead to the production of hazes.  \citet{Morley2015} find that as planet temperatures decrease towards 500~K, soot-precursor production peaks and does so at higher altitudes, thus muting the size of spectroscopic features and reducing H\sb{2}O~--~J index values.  They also predict soot precursor production to slow below $\sim$500~K; therefore, more temperate exoplanet atmospheres could return to clearer skies and the H\sb{2}O~--~J index may rebound to values greater than unity.  However, the presence of water ice clouds for objects below $\sim$350~K may again mute spectral features \citep{Sudarsky2000,Morley2014}.

\subsection{Surface Gravity Dependence}
\label{sec:logg}

\begin{figure}[tb]
\centering
\includegraphics[width=1.0\linewidth]{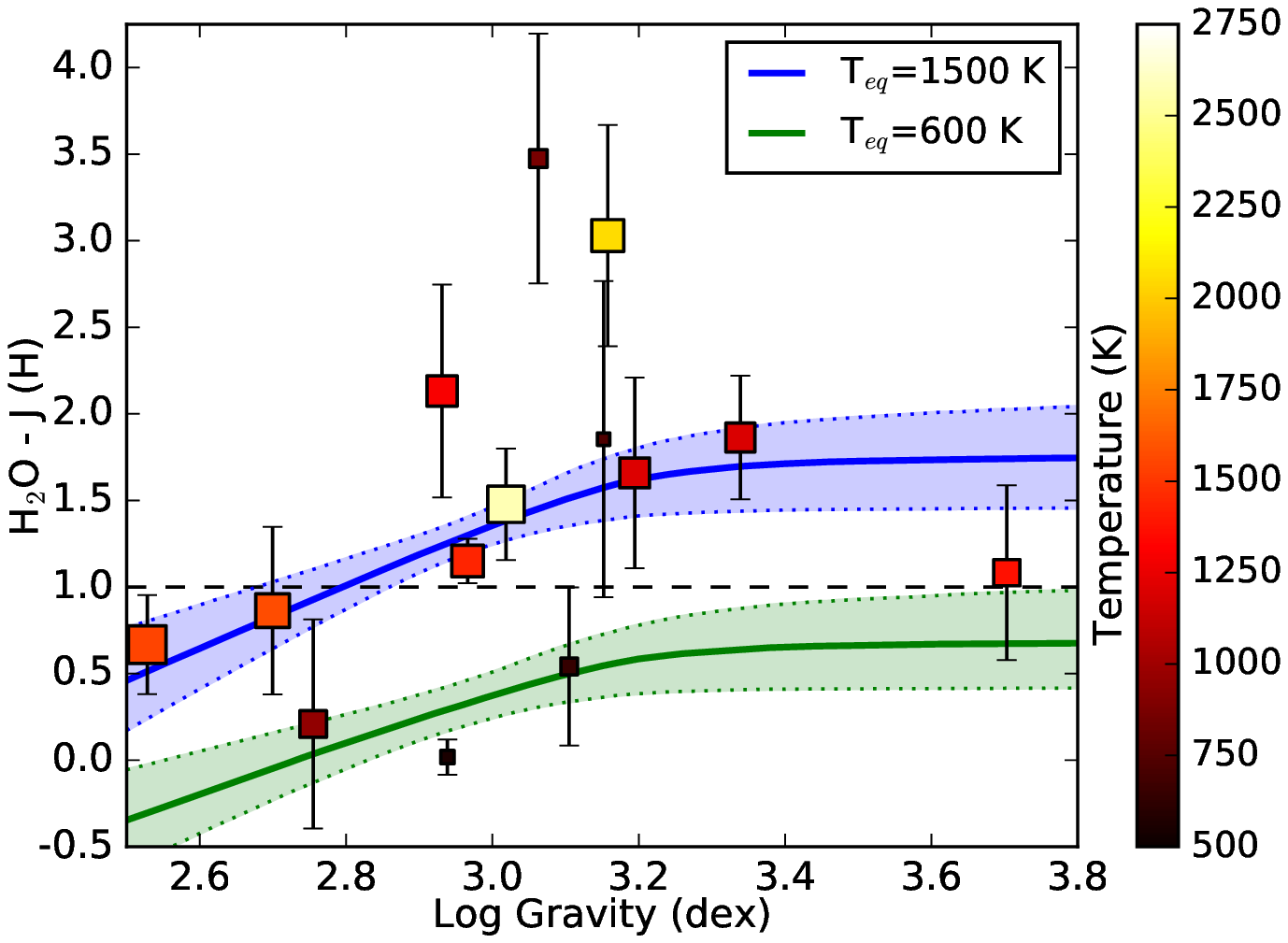}
\includegraphics[width=1.0\linewidth]{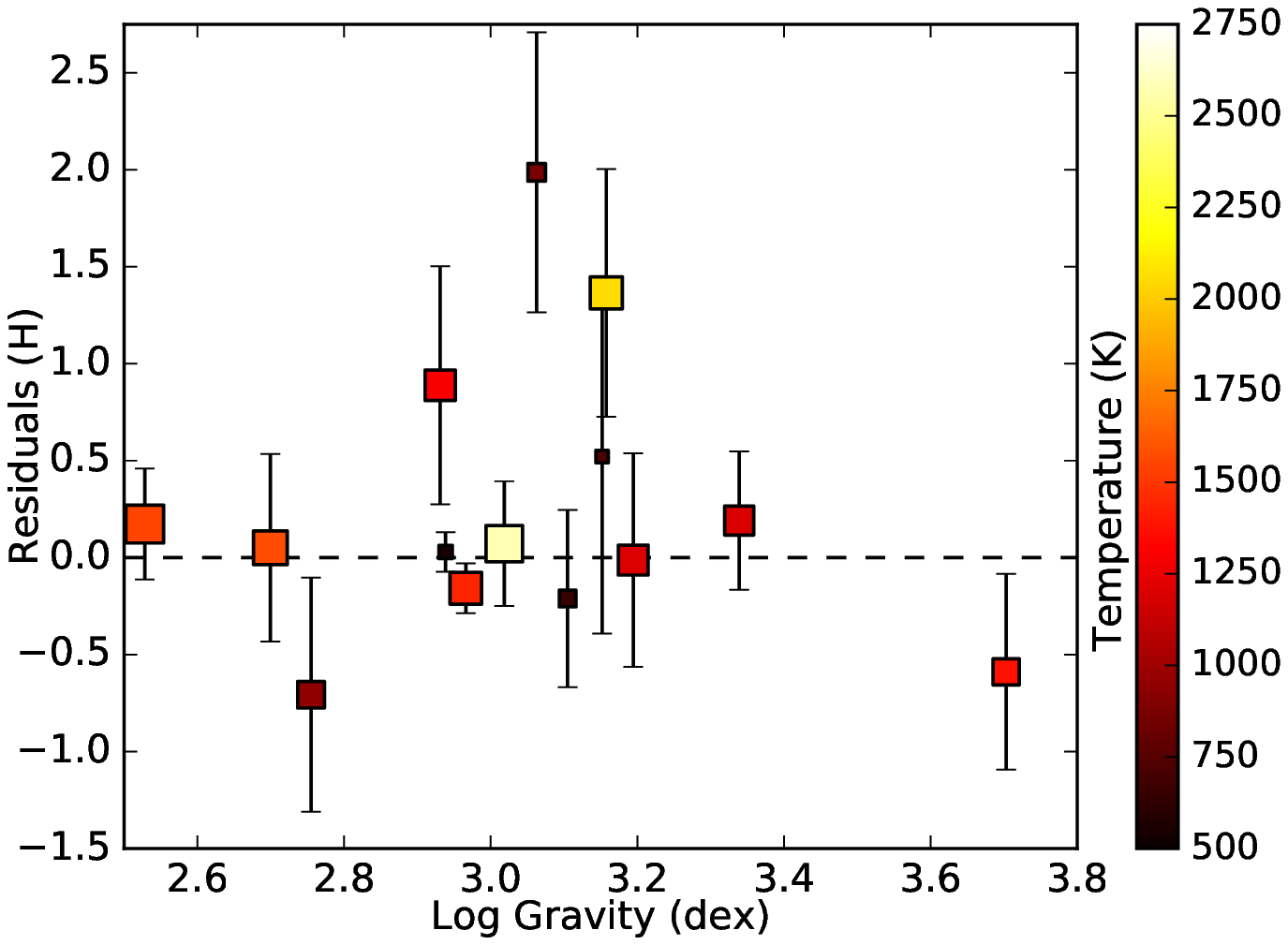}
\caption{\label{fig:loggDep}{
H\sb{2}O~--~J index dependence on surface gravity.  The symbol size is proportional to planet radius and symbol color is proportional to equilibrium temperature.  The top panel depicts the water feature strength with 1$\sigma$ uncertainties.  The blue and green lines with colored 1$\sigma$ uncertainty regions are representative models at two different temperatures ($T$\sb{eq} = 1500 and 600~K, respectively).  Curvature in the median model and uncertainties results from allowing the knot value to vary in our fits.  The bottom panel shows no discernible correlation between the residuals and surface gravity.}}
\end{figure}

Relative to planet temperature, we detect a weak correlation between the H\sb{2}O~--~J index and surface gravity.  In the top panel of Figure \ref{fig:loggDep}, we plot the calculated H\sb{2}O~--~J index values (with uncertainties) as a function of surface gravity for our entire sample population.  The correlation between these two parameters is difficult to assess without accounting for the temperature dependence.  Nevertheless, the two representative models at fixed equilibrium temperatures clearly show a decreasing trend towards lower-gravity planets.  The significance of this trend is 2.9$\sigma$.  The addition of a precise measurement from a low-surface-gravity exoplanet would improve this significance while providing another data point within a poorly sampled region of phase space.  The bottom panel of Figure \ref{fig:loggDep} displays the same H\sb{2}O~--~J index residuals as Figure \ref{fig:TeqDep}, except plotted as a function of surface gravity.

One possible explanation for the gradual decrease in water feature size with decreasing surface gravity is an increase in vertical mixing efficiency.  This is because as surface gravity decreases, the eddy diffusion coefficient ($K$\sb{zz}) increases \citep{Barman2011}.  As a result, enhanced vertical mixing could loft more soot precursors to higher altitudes within the atmosphere where they would effectively reduce the size of spectroscopic features.  However, increased photochemical production does not always occur under all atmospheric conditions and requires that soot precursors can initially form within the atmosphere, which is most likely for planets with temperatures between 500 -- 800~K \citep{Morley2013}.  Our sample of low-gravity exoplanets are all hotter than 800~K on their daysides, but their nightside temperatures are likely cool enough to favor the production of methane (and higher hydrocarbons) in thermochemcial equilibrium.  It remains to be seen whether high-altitude hazes can be produced on a low-gravity planet's nightside and whether they can remain optically thick at the planet terminator as they are transported to the dayside.

Since photochemical hazes are not favored at higher temperatures, another possible explanation is the presence of obscuring clouds.  Low-gravity environments can hamper particulate sedimentation, resulting in high-altitude cloud formation.  As an example, \citet{Demory2013}, \citet{Hu2015}, and \citet{Shporer2015} report on three Kepler planets (Kepler-7b, Kepler-12b, and Kepler-42b) with asymmetric optical-light phase curves.  They attribute the westward shift in peak brightness (relative to the substellar point) to the formation of reflective clouds.  Interestingly, the lowest-gravity planets (Kepler-7b and Kepler-12b, which have $\log g = 2.70$ and 2.57~dex, respectively) have the strongest asymmetry in their phase curves.  Silicates such as enstatite (MgSiO\sb{3}) and forsterite (Mg\sb{2}SiO\sb{4}) are both highly reflective and condense at high altitudes ($P$ < 0.01 bar) between 1,500 -- 2,000~K \citep{Morley2015}.  Thus, in addition to the Kepler planets, patchy silicate clouds can naturally explain the low H\sb{2}O~--~J index values reported here for WASP-17b and WASP-31b.

\section{FUTURE WORK}
\label{sec:future}

In this investigation into correlations between water feature size and planet physical properties, we find that the H\sb{2}O~--~J index is strongly correlated ($r=0.96$, p-value = 0.092) with planet temperature when $T$\sb{eq}$<750^{+90}_{-60}$~K and moderately correlated ($r=0.61$, p-value = 0.053) with surface gravity for planets with $\log g < 3.2^{+0.3}_{-0.2}$~dex.  However, we need additional, high-precision measurements from planets with key physical attributes to refine these thresholds and develop a more complete understanding of these transition regions.  In particular, the low-gravity regime is poorly sampled and lacks a high-precision constraint to definitively confirm a surface gravity dependence.  Additionally, {\em HST}/WFC3 measurements of a 500 -- 800~K Jupiter-size exoplanet orbiting an F, G, or K star are needed to definitively rule out links between water feature size and planet radius or host star type.

Future studies, particularly those with {\em JWST}, will be able to utilize additional indices over an expanded wavelength range to further subdivide these two classes of exoplanets and examine additional factors that might affect the size of spectroscopic features (such as patchy cloud cover and metallicity).  Until then, these results can serve as a guide for the selection of future targets while enhancing our understanding of past measurements.

\acknowledgments

I appreciate the helpful feedback from Jacob Bean, Megan Bedell, Diana Dragomir, and Laura Kreidberg, as well as the thoughtful suggestions from the anonymous referee.  I thank Michael Line for supplying model transmission spectra.  I also acknowledge the contributors to SciPy, Matplotlib, and the Python Programming Language, the free and open-source community, and the NASA Astrophysics Data System for software and services.   
This work was performed under contract with the Jet Propulsion Laboratory (JPL) funded by NASA through the Sagan Fellowship Program executed by the NASA Exoplanet Science Institute.
\\


\end{document}